\documentclass{iopart}
\usepackage{iopams}
\usepackage{graphicx}

\newcommand{\ok}[1]{\left( #1 \right)}

\begin{document}

\title{Correlation Effects in  Side-Coupled  Quantum Dots}

\author{R \v{Z}itko$^1$ and J Bon\v{c}a$^{1,2}$}

\address{$^1$J. Stefan Institute, Jamova 39, SI-1000 Ljubljana, Slovenia}
\address{$^2$Department  of Physics,
FMF, University of Ljubljana, Jadranska 19, SI-1000 Ljubljana, Slovenia}
\ead{\mailto{rok.zitko@ijs.si}, \mailto{janez.bonca@ijs.si}}

\date{\today}

\begin{abstract}
Using Wilson's numerical renormalization group (NRG) technique we
compute zero-bias conductance and various correlation functions of a
double quantum dot (DQD) system. We present different regimes within a
phase diagram of the DQD system. By introducing a negative Hubbard U
on one of the quantum dots, we simulate the effect of electron-phonon
coupling and explore the properties of the coexisting spin and charge
Kondo state. In a triple quantum dot (TQD) system a multi-stage Kondo
effect appears where localized moments on quantum dots are screened
successively at exponentially distinct Kondo temperatures.
\end{abstract}


\bibliographystyle{unsrt}

\section{Introduction}

The development of nanotechnology has stimulated studies of transport
through coupled quantum dots systems, where at very low temperatures
Kondo physics as well as magnetic interactions play an important
role. A double-dot system represents the simplest generalization of a
single-dot system which has been extensively studied in the
past. Recent experiments demonstrate that similar realistic devices
can be constructed \cite{dqd-expr1,dqd-expr2,dqd-expr3,pcdqd}, which
enables direct experimental investigations of the competition between
the Kondo effect and the exchange interaction between localized
moments on the dots. One manifestation of this competition is a two
stage Kondo effect that has recently been predicted in multilevel
quantum dot systems with explicit exchange interaction coupled to one
or two conduction channels \cite{qptmultilevel, hofstetter2004}.
Experimentally, it manifests itself as a sharp drop in the conductance
vs. gate voltage $G(V_G)$ \cite{two_stage} or as non-monotonic
dependence of the differential conductance vs. drain-source voltage
$dI/dV_{ds}(V_{ds})$ \cite{granger2005}.

Another line of research of correlation effects in quantum dots
concerns the influence of localized phonon modes on electron transport
through quantum dots and molecules. The influence of a strong,
localized electron-phonon interaction \cite{mravlje} can be simulated
by the introduction of an effective negative-$U$ Anderson model. In
such system the {\it charge Kondo effect} leads to screening of charge
fluctuations on the localized impurity by low-frequency pair
fluctuations in the leads \cite{coleman92}. It has recently been shown
that in the case of large electron-phonon coupling the original
single-impurity Anderson model, coupled to local phonon degrees of
freedom, can be mapped onto an anisotropic Kondo model
\cite{cornaglia2004}. Increasing the electron-phonon coupling leads to
a suppression of the conductance plateau width
\cite{cornaglia2004,mravlje} as a function of the gate voltage.

\begin{figure}[htb]
\centering
\includegraphics[totalheight=2cm]{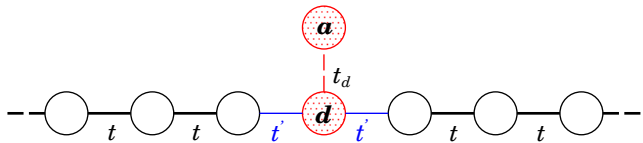}
\caption{(Color online) Side-coupled configuration of quantum dots}
\label{figa1}
\end{figure}

We study a double quantum dot (DQD) in a side-coupled configuration
(figure~\ref{figa1}), connected to a single conduction-electron
channel. Similar systems have been studied using various analytical
techniques such as the non-crossing approximation \cite{suppression},
embedding technique \cite{topology} or slave-boson mean field theory
\cite{kang,sidedouble}, and recently also using more accurate
numerical renormalization group (NRG) calculations \cite{corn}.

In this work we study three distinct cases. In the case, already
studied in detail in Refs~\cite{corn,zitko2006a,zitko2006b}, when
effective Coulomb interactions on DQD are positive and the intra-dot
overlap is large, we find wide regimes of enhanced conductance as a
function of gate-voltage at low temperatures due to the Kondo
effect. Regimes of enhanced conductance are separated by regimes where
localized spins on DQD are antiferromagnetically (AFM) coupled.  In
the limit when the dot $a$ is only weakly coupled, the system enters
the "two stage" Kondo regime \cite{vojta, corn}, where we again find a
wide regime of enhanced conductivity under the condition that the
high- and the low- Kondo temperatures ($T_K$ and $T_K^0$ respectively)
are well separated and the temperature of the system $T$ is in the
interval $T_K^0 \ll T \ll T_K$.

In the second case we explore the coexistence of the spin and charge
Kondo state with different Kondo temperatures that are realized when
there is an effective attractive interaction on one of the quantum
dots. We show that spin and charge susceptibilities are screened at the
corresponding spin and charge Kondo temperatures. Such a spin-charge
Kondo state can be realized when two molecules, one with a strong
phonon mode and the other with a localized orbital where electrons
experience strong Coulomb interaction, would be coupled to metallic
leads.

Finally, in the case of three side-coupled quantum dots we demonstrate
that a multi-stage Kondo effect is realized provided that exchange
couplings between successive quantum dots are smaller than the
corresponding Kondo temperatures.

\section{Model and method}

The Hamiltonian that we study reads
\begin{eqnarray}
H &=
\delta_d (n_d-1) + \delta_a (n_a-1)
- t_d \sum_\sigma \left( d^\dag_\sigma a_\sigma + a^\dag_\sigma d_\sigma
\right) \\
&+ \frac{U_d}{2} (n_d-1)^2 + \frac{U_a}{2} (n_a-1)^2 \\
&+ \sum_{k\sigma} \epsilon_k c^\dag_{k\sigma} c_{k\sigma} +
\sum_{k\sigma} V_d(k) \left( c^\dag_{k\sigma} d_{\sigma} +
d^\dag_\sigma c_{k\sigma} \right), \label{ham}
\end{eqnarray}
where $n_d=\sum_\sigma d^\dag_\sigma d_\sigma$ and $n_a=\sum_\sigma
a^\dag_\sigma a_\sigma$. Operators $d^\dag_\sigma$ and $a^\dag_\sigma$
are creation operators for an electron with spin $\sigma$ on site $d$
or $a$.  On-site energies of the dots are defined by
$\epsilon=\delta-U/2$.  For simplicity, we choose the on-site energies
and Coulomb interactions to be equal on both dots,
$\delta_a=\delta_d=\delta$.  Coupling between the dots is described by
the inter-dot tunneling coupling $t_d$. Dot $d$ couples to both leads
with equal hopping $t'$.  As it couples only to symmetric combinations
of the states from the left and the right lead, we have used a unitary
transformation \cite{glazmanraikh} to describe the system as a variety
of the single-channel, two-impurity Anderson model.  Operator
$c_{k\sigma}^\dag$ creates a conduction band electron with momentum
$k$, spin $\sigma$ and energy $\epsilon_k=-D \cos{k}$, where $D=2t$ is
the half-bandwidth.  The momentum-dependent hybridization function is
$V_d(k)=-(2/\sqrt{N+1})\, t' \sin{k}$, where $N$ in the normalization
factor is the number of conduction band states.

We use Meir-Wingreen's formula for conductance in the case of
proportionate coupling \cite{meirwingreen} which
is known to apply under very general conditions
(for example, the system need not be in a Fermi-liquid ground state)
with spectral functions obtained using the NRG technique
\cite{wilson, magnetocosti, sia1, hofstetter}.
At zero temperature, the conductance is
\begin{equation}
G=G_0 \pi \Gamma A_d(0),
\label{adom}
\end{equation}
where $G_0=2e^2/h$, $A_d(\omega)$ is the local density of states of
electrons on site $d$ and $\Gamma/D=(t'/t)^2$.

We use the density-matrix \cite{hofstetter,zitko2006a} version of the
standard NRG method \cite{wilson}. In all our NRG calculations we took
into account the rotational invariance in the spin space as well as in
the isospin space, {\it i.e.} $SU(2)_\mathrm{spin} \otimes
SU(2)_\mathrm{isospin}$.

\section{ Phase diagram of the DQD system}

\begin{figure}[htbp] 
\centering
\includegraphics[width=8cm]{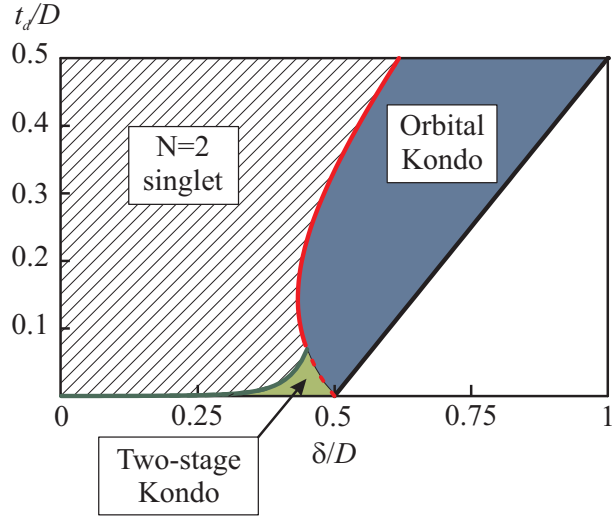}
\caption{ Phase diagram of a DQD for $U/D=1$, obtained using
analytical estimates as given in the text. Grey area represents the
region of the Kondo regime where $S\sim 1/2$, $\langle n\rangle \sim
1$, and $G/G_0\sim 1$. In the shaded area, called spin - singlet
regime, where $S\sim 0$ and $\langle n\rangle \sim 2$, spin-spin
correlation function is enhanced, {\it i.e.}  $\langle {\bf S}_a \cdot
{\bf S}_d \rangle \lesssim -0.5$ and $G/G_0\sim 0$. The two-stage
Kondo regime is explained further in the text.  }
\label{fig_pd}
\end{figure}

As a part of the introduction, we first review some findings on the
model with repulsive Hubbard interaction $U_d=U_a=U>0$ as already
given in Refs.~\cite{zitko2006a,zitko2006b}. We first present the
phase diagram of the DQD system as depicted in
figure~\ref{fig_pd}. Regime of high conductance is represented by the
dark-gray area. High conductance is a consequence of the Kondo regime
where $\langle n \rangle \sim 1$.  Full lines defining the Kondo
regime are given by the following analytical expressions: $\delta_1=
t_d (2\sqrt{1+(U/4t_d)^2}-1)$ and $\delta_2=(U/2+t_d)$. A spin-singlet
regime forms in the strong coupling regime as soon as $\langle n
\rangle \sim 2$, while in the weak coupling regime, {\it i.e.}  for
$t_d/D\lesssim 1$ the spin-singlet state persists down to the
following condition $J_{\mathrm{eff}} = T_K$. In the opposite case,
when $J_{\mathrm{eff}}<T_K$, exponentially small scale, the two-stage
Kondo regime appears for $T<T_{K}^0$. At low temperatures, the
conductance is enhanced when $T_K^0\lesssim T \lesssim
J_{\mathrm{eff}}$.

We now explore in detail the limit when $t_d\to 0$.  In this case one
naively expects to obtain essentially identical results as in the
single-dot case. It turns out, however, that a non-zero inter-dot
coupling, $t_d \not = 0$, represents a relevant perturbation to the
system in the limit $T\to 0$.  A clear physical picture is obtained by
comparing the two relevant energy scales: the exchange interaction
$J_{\mathrm{eff}}$ and the single impurity Kondo temperature. In the
region defined by $J_{\mathrm{eff}}<T_K$ and $\delta<U/2$, the system
enters the two-stage Kondo regime. In this regime, a gap of width
$T_K^0$ opens in the density of states $A_d(\omega)$ \cite{corn} which
consequently leads to a decrease of the conductance when the
temperature is below $T_K^0$.  The lower Kondo temperature $T_K^0$
depends exponentially on $T_K$ as \cite{corn, vojta}
\begin{equation}
T_K^0 \sim T_K \exp(-\alpha T_K/J_{\mathrm{eff}}).
\end{equation}
When by decreasing $\delta$ or by increasing $t_d$, the exchange
interaction overcomes $T_K$, {\it i.e.} $J_{\mathrm{eff}}>T_K$, the
gap in $A_d(\omega)$ is given by $J_{\mathrm{eff}}$. In this regime
the two local moments on DQD form a spin-singlet.

\begin{figure}[htbp]
\centering
\includegraphics[width=6.5cm,angle=0]{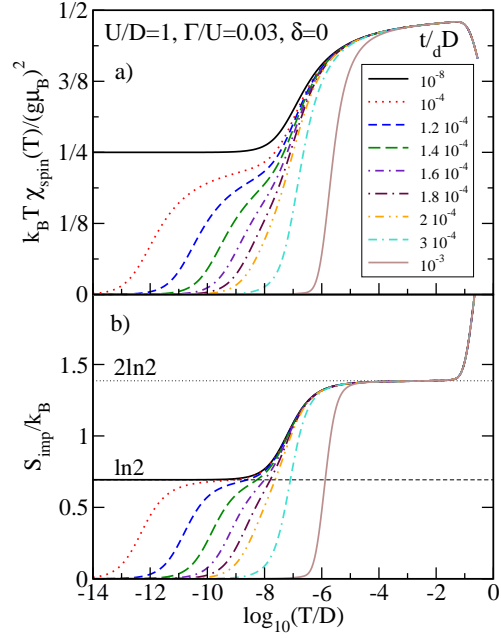}
\caption{ Temperature-dependent spin-susceptibility a), and entropy b)
  for various values of $t_d/D$, computed at $\delta=0$,
  $\Gamma/D=0.03, U/D=1$. See also Ref.~\cite{corn}.}
\label{fig_chis0}
\end{figure}

The two-stage Kondo regime can be clearly observed in
Fig~\ref{fig_chis0}a  where we show the impurity
part of the spin susceptibility $\chi_{\mathrm{s}}$
\begin{equation}
\chi_{\mathrm{s}}(T)=\frac{\ok{g\mu_B}^2}{k_B T} \ok{ \langle S_z^2 \rangle
- \langle S_z^2 \rangle_0 }.
\end{equation}
The first expectation value in this expression refers to the system
with the double quantum dot, while the second refers to the system
without the dots.  The two-stage Kondo effect manifests as two
successive decreases of the susceptibility, first at $T \sim T_K\sim
10^{-7}$ followed by the second at $T \sim T_K^0$ \cite{corn}. In this
case, the two-stage Kondo effect occurs for $t_d/D \lesssim 1.6\
10^{-4}$.

The same effect can also be seen in the temperature-dependent impurity
contribution to the entropy $S_\mathrm{}(T)$, defined via
\begin{equation}
S_\mathrm{}(T) = \frac{\left( E-F \right)}{T}
- \frac{\left( E-F \right)_0}{T},
\end{equation}
where $E = \langle H \rangle = \mathrm{Tr} \left( H e^{-H/(k_B T)}
\right)$ and $F = -k_B T \ln \mathrm{Tr} \left( e^{-H/k_B T} \right)$.
At $T \sim T_K$ the entropy first decreases from $2\ln 2$ to $\ln 2$,
see Fig~\ref{fig_chis0}b. This drop indicates the Kondo screening of
the local moment on the dot-$d$ while the moment on the dot-$a$
remains unscreened and nearly free. The second drop takes place at $T
\sim T_K^0$.

\begin{figure}[htbp]
\centering
\includegraphics[width=6.5cm,angle=0]{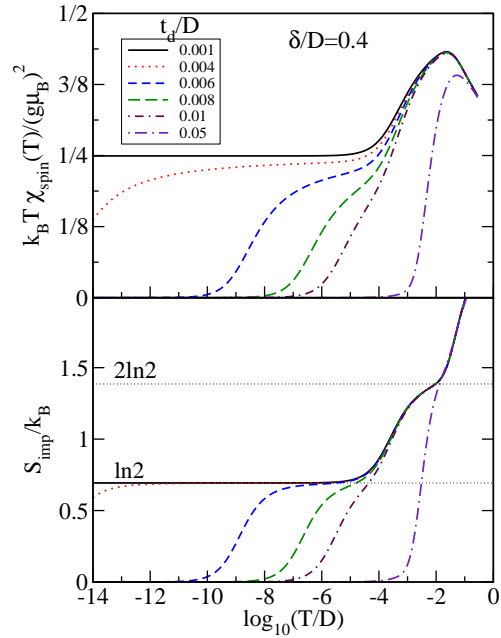}
\caption{  The same as in Fig~\ref{fig_chis0}, except for  $\delta=0.4$.}
\label{fig_chis1}
\end{figure}

With increasing $\delta$, $T_K$ increases, and the condition for the
two-stage Kondo effect is satisfied at increasingly larger $t_d/D$. In
Fig~\ref{fig_chis1} we present $\chi_\mathrm{s}(T)$ and
$S_\mathrm{}(T)$ calculated at $\delta/D=0.4$. The two-stage Kondo
effect sets in at substantially larger value of $t_d/D$ than in the
$\delta=0$ case, {\it i.e.} at $t_d/D\sim 0.01$.

\begin{figure}[htbp]
\centering
\includegraphics[width=6.5cm,angle=0]{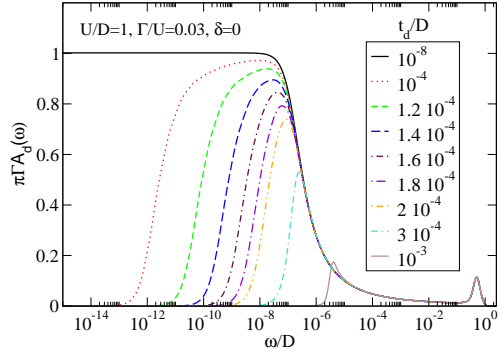}
\caption{ Normalized spectral function $ A_d(\omega)$ of DQD
  system. System parameters are identical to those used in
  Fig~\ref{fig_chis0}. See also Ref.~\cite{corn}.}
\label{fig_adom}
\end{figure}

In Fig~\ref{fig_adom} we explore how the two Kondo temperatures
reflect in the shape of the frequency dependent spectral function
$A_d(\omega)$, computed on the dot $d$. At $\omega\sim U/2$ all curves
display a charge-transfer peak. At lower frequencies spectral
functions again increase around $\omega\sim T_K$, and then decrease
around $\omega\sim T_K^0$. All calculations were done at the effective
temperature $T/D \sim 10^{-16}$, which is lower than $T_K^0$ for all
$t_d$, except for $t_d/D \sim 10^{-8}$. Taking into account the
Meir-Wingreen formula \cite{meirwingreen} for the zero-bias
conductance, equation~(\ref{adom}), we realise that in the limit $T\to
0$, $G/G_0\to 0$ for any finite $t_d/D$. At small but finite $T$, such
that $T_K^0 \lesssim T \ll T_K$, the gap in $ A_d(\omega=0)$ closes
and the conductance approaches its unitary limit $G/G_0\sim 1$.

\section{Spin versus charge Kondo effect}

In this section we investigate a system with $U_d=-U_a=U>0$. The
negative $U$ on dot-$a$ simulates the effect of strong electron-phonon
interaction. Beside the impurity spin susceptibility and entropy we
also compute the normalized charge susceptibility, defined as
\begin{equation} 
\chi_{\mathrm{c}}(T)=\frac{1}{ k_B T} \left( \langle I_z^2 \rangle
- \langle I_z^2 \rangle_0 \right),
\end{equation}
where $I^z$ is the $z$-component of the total isospin of the system.
The isospin operators are given by 
\begin{equation}
{\bf I}_i = \sum_{\alpha\alpha'}\eta^\dag_{i,\alpha}
\boldsymbol{\sigma}_{\alpha\alpha'} \eta_{i,\alpha'},
\end{equation}
where the Nambu spinor $\eta^\dag_i$ on the impurity orbitals are defined by
\begin{equation}
\eta^\dag_a = 
\left(\begin{array}{c}
a^\dag_{\uparrow} \\
- a_{\downarrow}
\end{array}\right),
\qquad
\eta^\dag_d = \left(
\begin{array}{c}
d^\dag_{\uparrow} \\
d_{\downarrow}
\end{array}\right), 
\end{equation}
Isospin raising operator is defined as $I^+ = d^\dag_{\downarrow}
d^\dag_{\uparrow} - a^\dag_{\downarrow}a^\dag_{\uparrow}$. Similarly,
isospin operators can be defined on the Wilson's chain and the total
isospin operator can then be defined by a sum of isospin operators
over all orbitals of the problem (impurities and conduction band). For
$\delta=0$, both ${\bf I}^2$ and $I_z$ commute with $H$ and $I$ and
$I_z$ are additional good quantum numbers.

Since the system consists of two quantum dots, one with positive- and
the other with negative- $U$ we expect that the dot-$d$ would display
a spin Kondo effect where the local spin moment would be screened by
spin fluctuation in the lead, while dot-$a$ would display a charge
Kondo effect where charge degrees of freedom on dot-$a$ would be
screened by pair fluctuations in the leads. Due to different overlaps
with the leads we expect the spin Kondo temperature $T_{KS}$ to be
different from the charge Kondo temperature, $T_{KC}$.

\begin{figure}[htbp] 
\centering
\includegraphics[width=6.5cm,angle=0]{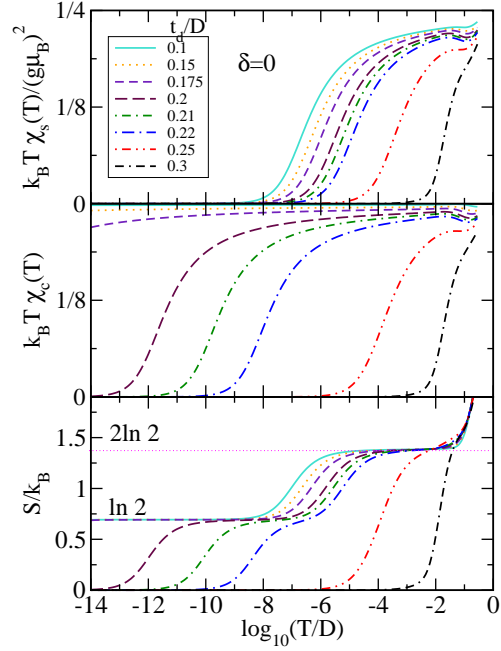}
\caption{ Temperature-dependent $\chi_{\mathrm{s}},\chi_{\mathrm{c}}$
and $S$ for various values of $t_d/D$, computed at $\delta=0$,
$\Gamma/D=0.03, U_d/D=-U_a/D=1$.}
\label{fig_unu}
\end{figure}

In figure~\ref{fig_unu} we present results of
$\chi_{\mathrm{s}}(T),\chi_{\mathrm{c}}(T)$ and $S(T)$ for various
inter-dot coupling strengths. At large $t_d/U\gtrsim 0.3$ both
susceptibilities as well as the entropy show a rapid drop with the
temperature, characteristic for a system where the ground state of the
DQD is a spin and isospin singlet ( state with no spin nor charge
fluctuations). This finding in agreement with the solution of the DQD
system in the atomic limit. For $t_d>\sqrt{3} U/6\sim 0.29U$, the
ground state is nondegenerate spin and isospin singlet ($S=0,I=0$)
with energy 
\begin{equation}
E_0({S=0,I=0})=-2t_d.
\end{equation}
In the opposite case, the ground state becomes four-fold degenerate
spin doublet ($S=1/2$), isospin doublet ($I=1/2$) with energy
\begin{equation}
E_0({S=1/2,I=1/2})=-\sqrt{t_d^2+U^2/4}.
\end{equation}
In the $t_d\to 0$ limit, the latter state is composed of a spin (up or
down) moment on the dot-$d$ and zero $(I_z=-1/2)$ or doubly occupied
$(I_z=+1/2)$ dot-$a$. All curves in figure~\ref{fig_unu} for
$t_d/D\lesssim 0.22$ show two separate Kondo screenings: spin Kondo
effect is visible in $\chi_{\mathrm{s}}(T)$ at $T_{KS}$ and charge
Kondo effect is visible in $\chi_{\mathrm{c}}(T)$ at
$T_{KC}<T_{KS}$. This is corroborated by the shape of $S(T)$ that
shows a plateau at $2\ln 2$ at $T> T_{KS}$, in agreement with the
four-fold degenerated $S=I=1/2$ ground state, a drop at $T\sim
T_{KS}$, followed by a plateau at $\ln 2$ due to the uncompensated
isospin moment (degenerate zero and doubly occupied states on the
dot-$a$), and finally a drop at $T\sim T_{KC}$. Decreasing $t_d/D$
affects $T_{KC}$ much more than $T_{KS}$. A simple explanation where
the effective hybridization of the dot-$a$ strongly depends on $t_d/U$
while hybridization of the dot-$d$ becomes $t_d$-independent for small
$t_d$ seems to be adequate.

\begin{figure}[htbp]
\centering
\includegraphics[width=10cm,angle=0]{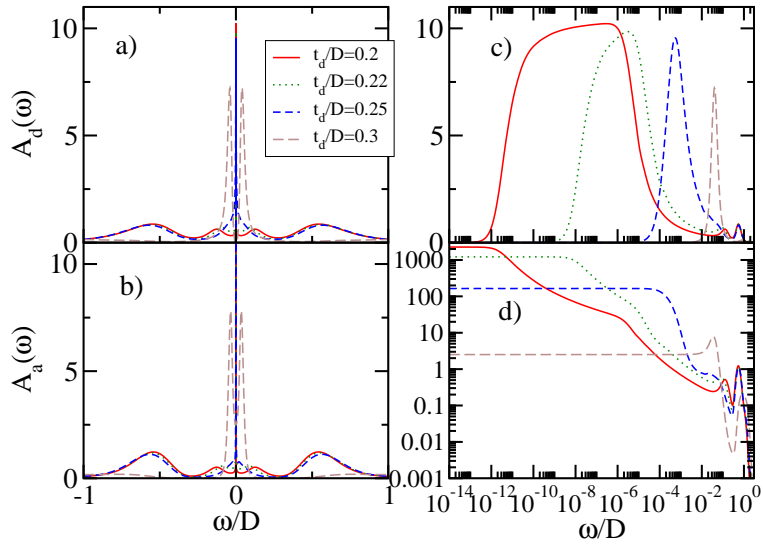}
\caption{ Spectral functions $ A_d(\omega)$ a) and c) $ A_a(\omega)$
  b) and d) of DQD
  system. System parameters are identical to those used in
  Fig~\ref{fig_unu}. }
\label{fig_adaom}
\end{figure}

In Fig.~\ref{fig_adaom} we present spectral functions of the DQD
system. In all figures there is a charge-transfer peak, located at
$\omega = \sqrt{t_d^2+U^2/4}$. There is an additional high-frequency
peak which is, for $t_d<\sqrt 3 U/6$, located at $\omega =
\sqrt{t_d^2+U^2/4} - 2t_d$. In this regime, {\it i.e.} for $t_d<\sqrt
3 U/6$, a sharp increase in $A_d(\omega)$ at $\omega\sim T_{KS}$
signals development of the spin Kondo peak on the dot-$d$, while at
lower frequency $\omega\sim T_{KC}$ a gap opens in $A_d(\omega)$. In
contrast, there is an additional increase of spectral weight in
$A_a(\omega)$ below this frequency. In the limit of $\omega\to 0$
there is an isospin Kondo peak in $A_a(\omega)$, mirrored by a gap of
width $\Delta\omega\sim 2T_{KC}$ within the spin Kondo peak in
$A_d(\omega)$.

\section{Three-stage Kondo model in a side-coupled TQD model}

\begin{figure}[htbp]
\centering
\includegraphics[width=8cm,clip]{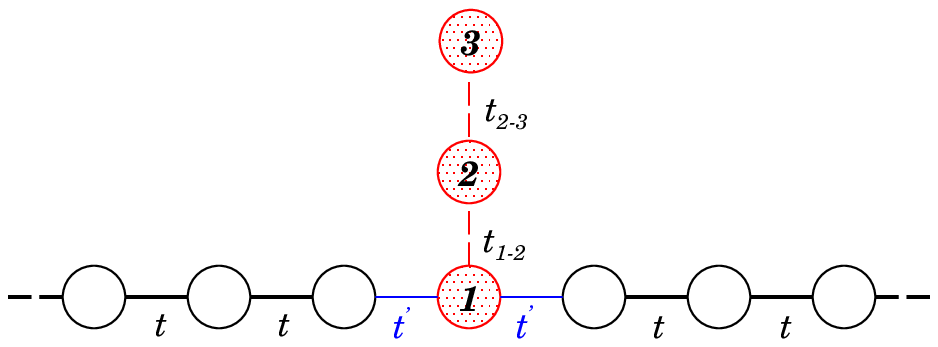}
\caption{ Side-coupled configuration of three quantum dots.}
\label{skica}
\end{figure}

\begin{figure}[htbp]
\centering
\includegraphics[width=8cm,clip]{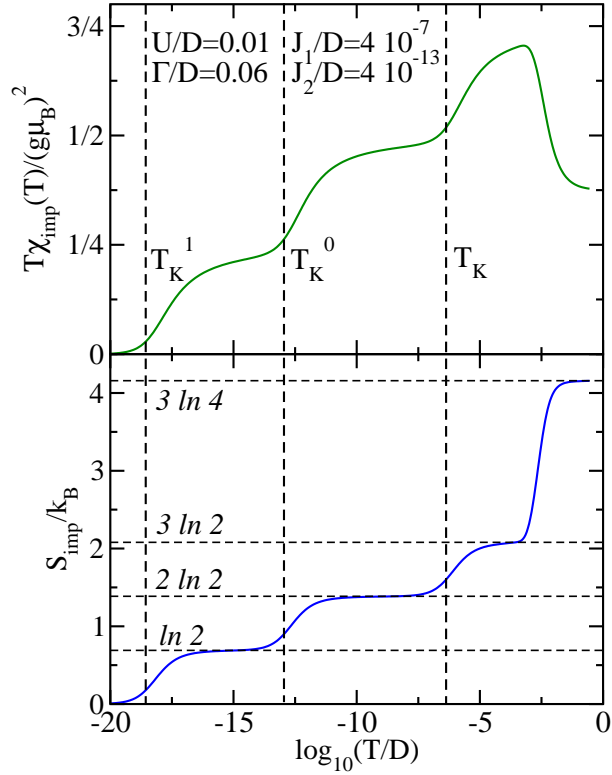}
\caption{Temperature-dependent susceptibility and entropy
for a side-coupled chain of three Kondo-like impurities.}
\label{fig_l}
\end{figure}

A natural question arises when studying the two-stage Kondo effect in
a side-coupled quantum dot: could one realise a multi-stage Kondo
effect in a series of side-coupled quantum dots.  We consider the
following model of three impurities, one coupled directly to a
continuum and two coupled in chain to the first one. Instead of
solving the more realistic three Anderson impurities model as depicted
in figure~\ref{skica}
we solve a simplified version of three Kondo impurities:
\begin{eqnarray}
H &= H_\mathrm{band} + \sum_i H_i + H_c + H_{123}\\
H_\mathrm{band} &= \sum_{k\sigma} \epsilon_k c^\dag_{k\sigma} c_{k\sigma}, \\
H_i &= \delta (n_i-1) + \frac{U}{2} (n_i-1)^2, \\
H_c &= \frac{1}{\sqrt{L}} \sum_{k\sigma} V_k \left( 
c^\dag_{k\sigma} d_{1\sigma} + d^\dag_{1\sigma} c_{k\sigma} \right), \\
H_{123} &= J_1 {\mathbf S}_1 \cdot {\mathbf S}_2 + J_2 {\mathbf S}_2 \cdot
{\mathbf S}_3,
\end{eqnarray}
where $J_1=-4t_{1,2}^2/U$ and $J_2=-4t_{2,3}^2/U$. In the case when
$J_1<T_K$, where $T_K$ is the single-impurity Kondo temperature of the
dot-1, and $J_2<T_K^0$ where $T_K^0$ is the second-stage Kondo
temperature, yet another low-energy scale 
\begin{equation}
T_K^1 \sim \exp (-\alpha T_K^0/J_2)
\end{equation}
appears in the problem. The system under the above conditions
undergoes three S=1/2 Kondo screening cross-overs, i.e. a three-stage
Kondo effect, as can be seen in Fig.~\ref{fig_l}. 

At $T\gtrsim T_K$, there is the local moment regime where $S/k_B\sim 3
\ln 2$ and the local moment approaches $T\chi_{\mathrm{imp}}/(g
\mu_B)^2\sim 3/4$, characteristic of three noninteracting local
spins. For temperatures below the first Kondo temperature, {\it i.e.}
$T\lesssim T_K$, the local moment on dot-1 couples into a singlet with
conducting electrons while spins on dots-2 and 3 remain uncoupled,
which gives $S/k_B \sim 2 \ln 2$ and $T\chi_{\mathrm{imp}}/(g
\mu_B)^2\sim 1/2 $. There is a second drop of $S$ and
$\chi_{\mathrm{imp}}$ at $T\lesssim T_K^0$ where the spin on the dot-2
becomes screened by the quasiparticles in the leads forming a
Fermi-liquid with the screened moment on the dot-1. The leftover of
the entropy $S/k_B\sim \ln 2$ as well as the local moment
$T\chi_{\mathrm{imp}}/(g \mu_B)^2\sim 1/2 $ in this temperature range
are due to unscreened spin on the dot-3. Finally, there is a total
screening of the local moment at $T\lesssim T_K^1$.

\begin{figure}[htbp]
\includegraphics[width=8cm,clip]{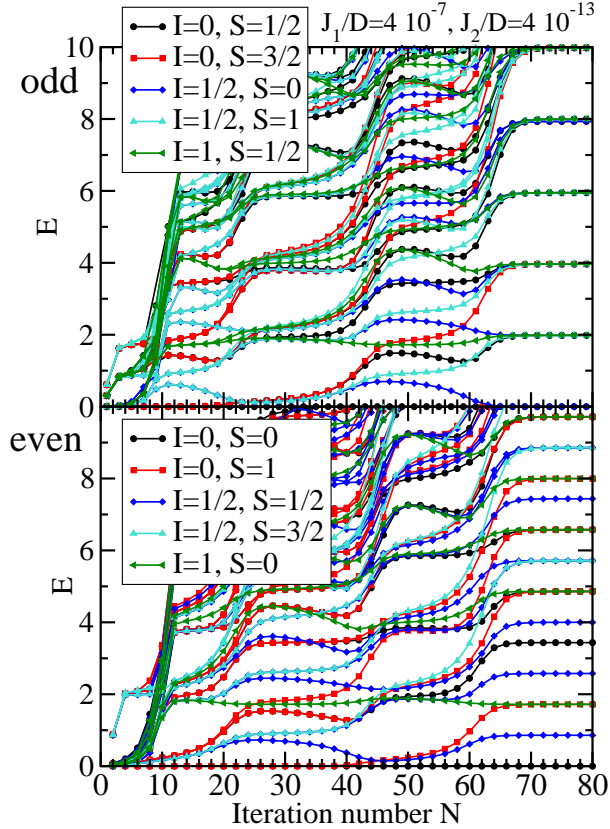}
\caption{Numerical renormalization group eigenvalue flows for odd and
even-length Wilson chains of the side-coupled triple quantum dot
system. The system exhibits three-stage Kondo screening at three
different Kondo temperatures. The levels are labeled by the quantum
numbers $I$ and $S$, the total isospin and total spin of the system.}
\label{fig_k}
\end{figure}

The numerical renormalization group eigenvalue flow is shown in
Fig.~\ref{fig_k}. Each Kondo screening induces an additional $\pi/2$
quasiparticle scattering phase shift while one ``free'' local moment
is simultaneously removed from the problem. Each $\pi/2$ phase shift
corresponds to a change of the boundary condition for the conduction
band electrons from periodic to anti-periodic (or vice-versa) or, from
another point of view, each Kondo screening removes one conduction
band site from the Wilson chain \cite{wilson}. As a consequence, we
observe an alternation between two sets of fixed point energies, each
set corresponding to one possible boundary condition. Note that
neither quantum numbers nor the degeneracy of the levels composing one
{\it energy} level remain the same after two such alternations: this
is due to the successive removal of the ``free'' local moments from
the problem. The degeneracy of the group of states having
(approximately) the same energy decreases after each Kondo cross-over;
in particular, the eigenvalues corresponding to high total spin are
pushed to higher energies. Of course, on the final strong-coupling
fixed point is stable.

\section{Conclusions}

We have explored different regimes of the side-coupled DQD. In the
case of positive Hubbard interactions when quantum dots are {\it
strongly coupled}, wide regions of enhanced, nearly unitary
conductance exist due to the underlying Kondo physics. Results were
summarised in a phase diagram of the DQD structure.  When DQD is
doubly occupied, conductance is zero due to formation of the
spin-singlet state which is effectively decoupled from the leads.
When quantum dots are {\it weakly coupled} unitary conductance exists
at finite temperatures when two electrons occupy DQD due to a
two-stage Kondo effect as long as the temperature of the system is
well below $T_K$ and above $T_K^0$.

It should be noted that if the signs of {\it both} electron-electron
interaction parameters $U_d$ and $U_a$ are inverted, we obtain an
equivalent problem, the only change being that the roles of spin and
isospin are permuted. If both quantum dots have negative $U$, we will
thus observe enhanced conductance due to the charge Kondo effect for
strong inter-dot tunneling coupling and suitable $\delta$, and
two-stage charge Kondo effect for weak inter-dot coupling.

We have explored a system where one of the dots is strongly coupled to
phonon degrees of freedom. This effect was simulated by the
introduction of an effective negative-$U$ interaction on one of the
dots, while $U$ on the other dot was kept positive. The main goal was
to explore the possibility of the coexistence of the spin and charge
Kondo effect. We found out that, indeed, spin and charge Kondo effect
can coexist. Even though our model contains two leads, due to mirror
symmetry the underlying problem falls into the class of the
single-channel Kondo problems. It is worth stressing that as a
consequence of the spin-charge separation in the leads, a single
channel is sufficient to completely screen two different moments -
spin as well as isospin.  For $U_d>0$, $U_a<0$ and small $t_d$, the
local moment (spin) on the directly coupled dot occurs at a higher
spin Kondo temperature, while the isospin degree of freedom on the
side-coupled dot is screened at much lower charge Kondo
temperature. For $U_d<0$ and $U_a>0$ the roles of the spin and charge
are simply reversed: there is a charge Kondo effect on the directly
coupled dot and spin Kondo effect on the side-coupled dot.

In the case of TQD, a multi-stage Kondo effect exists for the
appropriate choice of the exchange coupling constants. In the zero
temperature limit, the conductance approaches unitary limit in the
particle-hole symmetric case, since the TQD fixed point is identical
to the single-impurity fixed point with unitary conductance in the
particle-hole symmetric point. In fact, this conclusion can be
generalized in two ways. First of all, the fixed point and the
quasiparticle phase shift in the case of three impurities is the same
irrespective of the values of exchange constants. Depending on the
exchange constants, we can either have three-stage Kondo effect (weak
exchange coupling), S=1/2 Kondo effect for an antiferromagnetic chain
composed of the three impurities (both exchange constants large), or
two of the neighboring impurities couple into a local singlet, while
the spin on the third impurity is Kondo screened (if one of the
exchange constants is much larger than the other). The conductance is
unitary in all these cases. We can also generalize our findings for
the case of more than three impurities. In the case of N side-coupled
quantum dots we predict unitary zero temperature conductance in the
case of odd N and zero conductance for even N. For suitable exchange
constants, the screening can occur in several (N or less) stages and
the conductance as a function of the temperature will be
non-monotonous.

\ack
Authors acknowledge useful discussions with A. Ram\v sak and the
financial support of the SRA under grant P1-0044.

\bibliography{vsi}

\end{document}